**Title:** Electronic antiferromagnetic correlations in cuprates phase diagram

**Author:** Jiaxin Yin

**Affiliation:** Institute of Physics, Chinese Academy of Sciences, Beijing 100190, China

**Abstract**: Modern state-of-the-art techniques allow us to explore the magnetic and electronic structures of cuprates throughout the whole phase diagram, which defines the central questions on their emergent high temperature superconductivity. However, a simplified and unified description of the order parameters in these colorful phases together with their complex relations is still undetermined. Here we establish this phase diagram and the related structures based on recent experimental progresses with emphasizing several essential equations, and we try to understand them under the framework of antiferromagnetic (AF) correlation using the simplest mathematics. This model independent description of cuprates phase diagram gives both clues and constraints to a final microscopic theory of the cuprates superconductivity.

The advent of cuprate superconductors have stimulated numerous techniques producing well documented results in exploring the transport, thermodynamics, electrodynamics, interface conduction, flux quantization, charge dynamics and spin dynamics of these fascinating materials. The general phase diagram of cuprates that embraces all the emergent phenomena related to their high temperature superconductivity is, however, viewed with some differences [1-4].

Firstly, the agreement is well achieved in the antiferromagnetism (AFM) phase where it also starts as a Mott insulator, and all its related phenomena are consistent with the spin *1/2* Heisenberg Model [5]. AF ordering temperature is characterized by $J_z$ which is much smaller than $J_{ab}$, hence above this ordering temperature or doping, 2D AF correlation still exists but undergoes a quantum melting. Another common region of these phase diagrams is the dome shape superconducting phase. Recently there are more and more evidences [6-12] suggesting that its order parameter is the coherent pairing gap opening on Fermi arc, and the gap at the arc tip $\Delta_{SC}$ scales with $T_C$: $2\Delta_{SC}=4.3T_C$ (1). From neutron scattering measurements [13,14] it is found the spin gap ($\Delta_{spin}$) and spin resonance ($E_g$) also scale with $T_C$: $\Delta_{spin}=3.8T_C$ (2); $E_g=5.8T_C$ (3).

When doping outside the superconducting region the system comes to Fermi liquid region with small correlations, while in underdoped region [2] below $T_{cr}$, the magnetic correlation is rather strong with its spin dynamics can still be described by 2D Heisenberg model. In this phase there are relations: $3T^*=T_{cr}$ (4); $T_{cr}=J_{eff}$ (5); $\xi=1\sim2a$ (6), where $J_{eff}$ is effective exchange interaction and $T^*$ is pseudogap temperature and $\xi$ is the magnetic correlation length. This region is also identified as strange metal phase [1,3,4].

Below this phase lies the most exotic region as the pseudogap phase [15,16]: in k-space, pseudogap opens from the antinodal with gradually eating up the Fermi surface creating Fermi arc; in r-space, mesoscale pattern (checkboard or stripe) is formed with characteristic unit cell length of the pattern as: $L=4a$ (7). Size of pseudogap $\Delta^*$ [9,10,17,18] scales with $T^*$ with the same *d* wave ratio strongly indicating it's a pre-pairing state: $2\Delta^*=4.3T^*$ (8). Meanwhile, because pseudogap behaves as a depletion of density of state near the antinodal [9,19,20], it also competes with superconductivity. Thus the final explanation should capture both sides. Other important issues related to pseudogap phase such as QCP and time reversal symmetry broken will not be discussed in this paper.

The last region is the fluctuating superconductivity region as $T_0$. Its line shape is asymmetric dome like [21] with its order parameter $\Delta_0$ as the extrapolation of the nodal coherent gap to antinodal point [11,22] which scales with $T_0$: $2\Delta_0=4.3T_0$ (9).

These nine related equations are marked in magnetic structure and gap structure in fig.1. The magnetic structure (fig. 1a) in *q* space shows a universal hourglass shape, with lower dispersion having stronger intensity along antinodal direction and upper dispersion having stronger intensity along nodal direction and in between a resonant peak. Additionally, a spin gap opens below the lower dispersion. The gap structure (fig. 1b) is defined in *k* space, so it is important to introduce the Fermi surface topology. While transport measurement is consistent with the small Fermi pockets at the nodal with its area proportional to doping level *x*, ARPES mostly reveals the large hole Fermi surface but with its spectra weight proportional to *x* [23]. Now more data [24-26] indicates that in pseudogap state Fermi arc actually belongs to part of the small Fermi pocket. The *d* wave superconducting gap is defined on the Fermi arc with its maximum $\Delta_{SC}$ at arc tip and when extrapolating arc gap to the antinodal point we get the fluctuating pairing gap size ($\Delta_0$) and pseudogap ($\Delta^*$) is also defined at antinodal.

We first try to understand the magnetic structure. Following the discussion in ref. 27 that the low energy magnetic resonance is related to AF scattering between the hot spots which near the arc tip or as the crossings of the Fermi surface and magnetic Brillouin Zone (BZ) boundary (the dash lines in fig 2a) in the superconducting state, it naturally involves $2\Delta_{SC}$. We further argue that there is an excitation like a charge mode for the spins which has the same size as average $d$ wave gap $E_C=(\Delta_{SC})_{ave}=1.5T_C$. Thus: $E_g=E_C+2\Delta_{SC}=5.8T_C$ (3). Taking into account the sign difference in the $d$ wave pairing state, AF mediated low energy scattering mainly occurs between a hot spot and half of the arc on the other side. And spin gap is the excitation between a hot spot and a nodal point: $\Delta_{spin}=E_C+\Delta_{SC}=3.7T_C$ which is quite close to eq. (2). In fig. 2 the scattering differential $\delta$ is along antinodal direction which is consistent with its intensity anisotropy. Up magnetic dispersion that represents the reduced AF spin wave dispersion is also coupled to the electronic band structure around $J_{eff}$, and its intensity anisotropy can be understand as its stronger coupling to the nodal where the band has a wider width thus more scattering space.

Considering this high energy magnetic interaction is local in nature, it has a characteristic energy scale $J_{eff}$ and its real space magnetic unit cell has a characteristic length $\sqrt{2}a$. Eq. (5) and (6) are obvious from this perspective. To understand eq. (4) pseudogap has to be clarified first. There are three aspects of pseudogap [15,16]: 1. AF correlation (SDW): it behaves as a band folding respect to magnetic BZ, so energy shift is highest near antinodal outside magnetic BZ. 2. Real space mesoscale structure: mesoscale structure formation means localization of quasi-particles, which should begin with the lowest speed electrons from Van Hove singularity locating at antinodal. So pseudogap behaves as a sudden opened gap with depletion of antinodal states [9,19,20]. 3. Fermi surface nesting (CDW): Fermi surface at antinodal in none magnetic phase is rather straight, leading to quasi-nesting of parallel Fermi surfaces.

Moreover, these three aspects are strongly correlated and should be considered collectively. Effects of AF correlation can drive the band structure reconstruction creating flatter band near magnetic zoom boundary, which generates more low speed electrons whose localization will enhance the mesoscale structure. Localization of the density of states near antinodal will shift energy of these electrons to higher energy and enhance local magnetic scattering. Finally, the band structure in the SDW phase is also compatible with that in the CDW phase and vectors of the nested Fermi surface and mesoscale structure are both quite close to $2\pi/4a$. However, condition of any of the three aspects is not perfect satisfied in general situation, so SDW, mesoscale structures and CDW tend to be fluctuating and as a combination of the three the pseuodogap is always robust. Taking the correlation length of mesoscale structure to be $\xi^*=4a$ at $T^*$ and magnetic correlation length to be $\xi=\sqrt{2}a$ at $T_{cr}$, and assuming magnetic correlation length $\xi$ is linearly increasing with decreasing $T$ in the first order, then we get eq. (4).

The local AF correlation not only provides the original d wave pairing interaction as captured by (8) but also means the spin favors a certain direction as if there is an effective magnetic field which is heterogeneous for superconductivity. These remind us the critical field for a superconductor: $H_C(T)=H(0)(1-(T/T_C)^2)$. If $T_C$ is replaced by $T^*$ as the original pre-pairing temperature, and assuming an phenomenological internal magnetic field smaller than $H(0)$ as $H_{int}$, then the calculated $T$ is the lowest temperature that electrons get paired. And we take $T=T_o$ which represents the fluctuating pairing temperature, $T^*=H_{int}/H(0)=J$ ($J\sim 1/3J_{eff}$, $J$ goes from $1$ to $0$ as doping increases). Then we get: $T^*=J$ (10) and $T_o=J\sqrt{1-J}$ (11). The results are plotted in fig. 3 as the dot-line and dash-line. The pre-pairing and competing natures of AF correlations are both captured by eq. (11).

Another aspect of pseudogap state is localization of antinodal quasi-particles, hence only Fermi arc quasi-particles are coherent and superconducting gap only opens on Fermi arc ($\Delta_{SC}=L_{arc}\Delta_0$, $L_{arc}$ is the relative arc length). Because $T_0$ and $T_C$ scale with $\Delta_{SC}$ and $\Delta_o$ respectively with same d wave ratio in eq. (1) and (9), we can have $T_C=L_{arc}T_0$. If we treat Fermi arc as part of the Fermi pockets whose area is proportional to doping $x$, then arc length is proportional to $\sqrt{x}$. Since $x\sim 1-J$, we have: $L_{arc}=\alpha\sqrt{1-J}$, where α is related to Fermi surface topology of different materials and in the order of 1. Finally we get: $T_C=\alpha J(1-J)$ (12). This is plotted with $\alpha=1$ in fig. 3 as solid-line. As can be seen its parabolic shape is well reproduced. And this result is actually consistent with finding $Tc\sim x\Delta^*$ [10,28], considering $x\sim 1-J$, $T^*=J$ and $2\Delta^*=4.3T^*$ (8).

Through the above discussion, we can see that the magnetic and electronic structure of cuprate superconductors couple elegantly to produce colorful emergent phases. To express such elegancy, we have formulated several equations as a starting point and we are looking forward a complete microscopic theory built on these descriptions.

**Acknowledgements**

The work was financially supported by National Science Foundation and Ministry of Science and Technology of China.

**Figures**

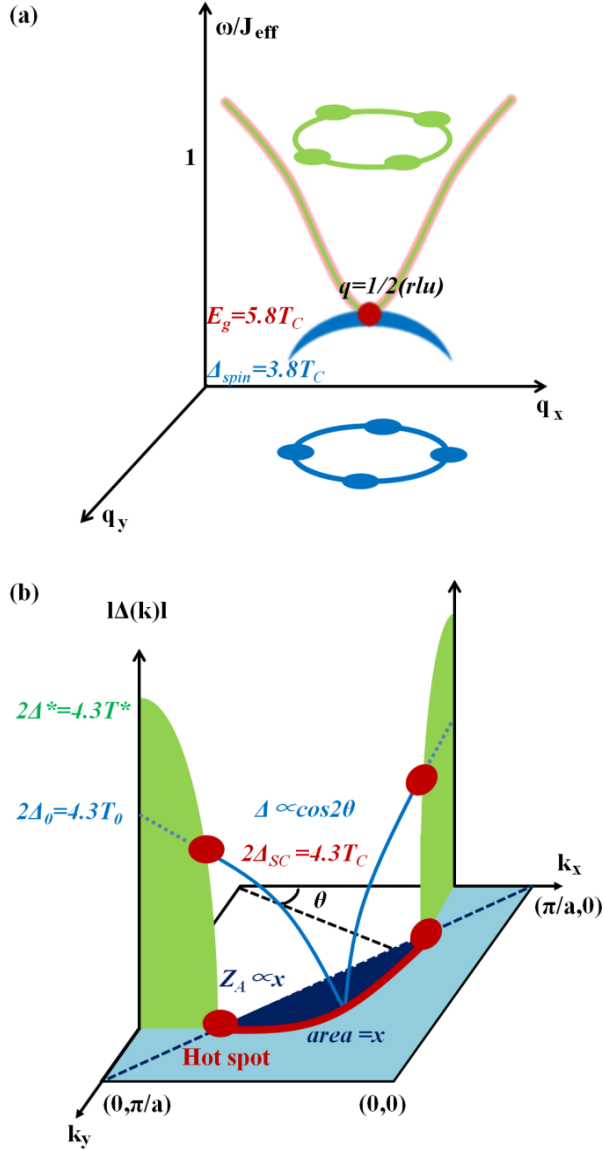

Fig. 1: (Colour on-line) (a) Hourglass like magnetic structure at $T \ll T_C$ based on neutron scattering [13]. Up dispersion (green curve) resembles the AF dispersion in parent compound with decreased band width and velocity which are both proportional to $J_{eff}$. Lower dispersion (blue curve) is argued as the coupling with low energy coherent electronic band dispersion [27]. The resonant peak ($E_g$) lies between this two dispersions and locate at $q_x=q_y=1/2$. A spin gap ($\Delta_{spin}$) is opened below lower dispersion. (b) Energy gap structure at $T \ll T_C$ [6-12,15-20]. The dash line is magnetic BZ. Red spots represent the hot spots as the crossings of the Fermi surface and magnetic BZ boundary. Green shade is pseudogap ($\Delta^*$) opening at antinodal. Blue curve is d wave superconducting gap. Red line in the Fermi surface is the Fermi arc which is characterized by coherent quasi-particle excitations, and its spectra weight is proportional to $x$. When considering Fermi arc as part of the Fermi pocket, dark blue area enclosed by arc and magnetic BZ is also proportional to $x$. Superconducting gap with its $d$ wave symmetry opens on the Fermi arc, with $\Delta_{SC}$ defined at the arc tip. When extrapolate blue curve to antinodal point we get fluctuating pairing gap size ($\Delta_0$). For both of the two structures, green color is related to the AF correlation, red color is related to superconducting phenomenon and blue color related to coherent quasi-particle states.

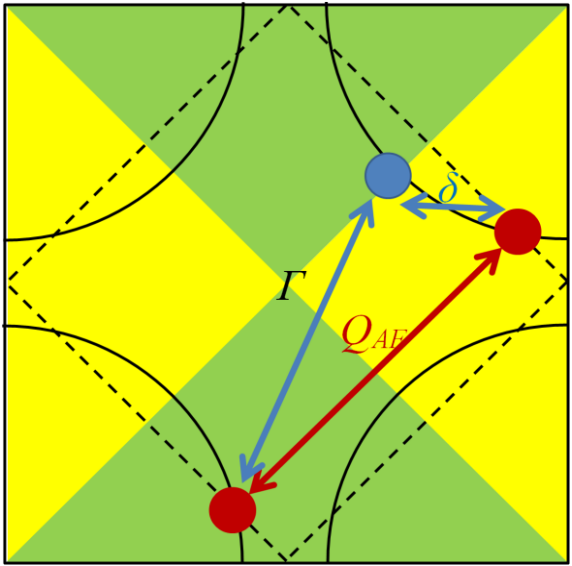

Fig. 2: (Colour on-line) The scattering channel related to resonant peak (red spot and line) and low incommensurate dispersion (blue lines) are marked around the Fermi surface. The yellow and green areas represent different superconducting phases with sign plus and minus respectively.

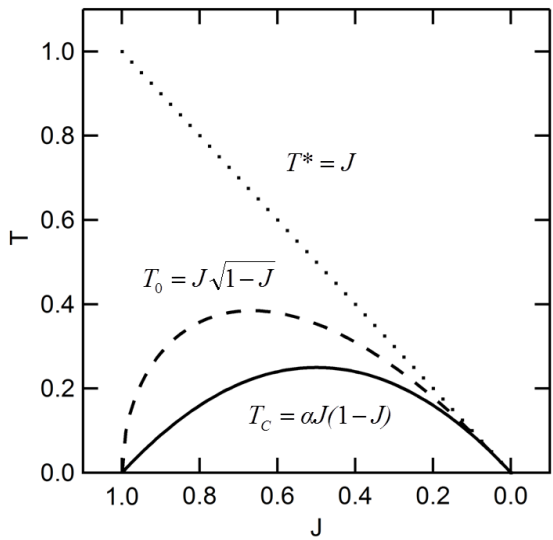

Fig.3: (Colour on-line) The evolution of critical temperature pseudogap $T^*$, fluctuating pairing $T_0$ and superconducting phases $T_{SC}$ with phenomenological magnetic exchange interaction $J$ as described by eq. (10) (11) (12) respectively.